# Large Scale In Silico Screening on Grid Infrastructures

Nicolas Jacq[1], Vincent Breton[1], Hsin-Yen Chen[2], Li-Yung Ho[2], Martin Hofmann[3], Hurng-Chun Lee[2], Yannick Legré[1], Simon C. Lin[2], Astrid Maaß[3], Emmanuel Medernach[1], Ivan Merelli[4], Luciano Milanesi[4], Giulio Rastelli[5], Matthieu Reichstadt[1], Jean Salzemann[1], Horst Schwichtenberg[4], Mahendrakar Sridhar[3], Vinod Kasam[1+3], Ying-Ta Wu[2], Marc Zimmermann[3]

[1] Laboratoire de Physique Corpusculaire, IN2P3 / UMR CNRS 6533,
24 avenue des Landais, 63177 AUBIERE, France
jacq@clermont.in2p3.fr
[2] Academia Sinica,
No. 128, Sec. 2, Academic Rd., NanKang, Taipei 115, Taiwan
ywu@gate.sinica.edu.tw
[3] Fraunhofer Institute for Algorithms and Scientific Computing (SCAI),
Schloss Birlinghoven, 53754, Sankt Augustin, Germany
martin.hofmann@scai.fraunhofer.de
[4] CNR-Institute for Biomedical Technologies,
Via Fratelli Cervi 93, 20090 Segrate (Milan), Italy
luciano.milanesi@itb.cnr.it
[5] Dipartimento di Scienze Farmaceutiche, Università di Modena e Reggio Emilia,
Via Campi 183, 41100 Modena, Italy
rastelli.giulio@unimore.it

**Abstract:** Large-scale grid infrastructures for in silico drug discovery open opportunities of particular interest to neglected and emerging diseases. In 2005 and 2006, we have been able to deploy large scale in silico docking within the framework of the WISDOM initiative against Malaria and Avian Flu requiring about 105 years of CPU on the EGEE, Auvergrid and TWGrid infrastructures. These achievements demonstrated the relevance of large-scale grid infrastructures for the virtual screening by molecular docking. This also allowed evaluating the performances of the grid infrastructures and to identify specific issues raised by large-scale deployment.

## 1 Introduction

In silico drug discovery is one of the most promising strategies to speed-up the drug development process [1]. Virtual screening is about selecting in silico the best candidate drugs acting on a given target protein [2]. Screening can be done in vitro but it is very expensive as there are now millions of chemicals that can be synthesized [3]. A reliable way of in silico screening could reduce the number of molecules required for in vitro and then in vivo testing from a few millions to a few hundreds.

In silico drug discovery should foster collaboration between public and private laboratories. It should also have an important societal impact by lowering the barrier to develop new drugs for rare and neglected diseases [4]. New drugs are needed for neglected diseases like Malaria where parasites keep developing resistance to existing drugs or Sleeping sickness for which no new drug has been developed for years. New drugs against Tuberculosis are also needed as the treatment now takes several months and is therefore hard to manage in developing countries.

However, in silico virtual screening requires intensive computing, of the order of a few TFlops per day to compute 1 million docking probabilities or for the molecular modelling of 1,000 ligands on one target protein. Access to very large computing resources is therefore needed for successful high throughput virtual screening [5]. Grids now provide such resources. A grid infrastructure such as EGEE [6] today provides access to more than 30,000 computers and is particularly suited to compute docking probabilities for millions of compounds. Docking is only the first step of virtual screening since the docking output data have to be further processed [7].

This paper will present the context and objectives of the two initiatives against Malaria and avian flu in chapters 2 and 3. In chapter 4, the grid infrastructures and the environments used for large-scale deployment are briefly described. Results and performance of grid environment are discussed in chapter 5. In chapter 6, we will also propose some perspectives for the coming years.

## 2   Large scale in silico docking against Malaria

Genomics research has opened new ways to find novel drugs to cure Malaria, vaccines to prevent Malaria, insecticides to kill infectious mosquitoes and strategies to prevent development of infectious sporozoites in the mosquito [8]. These studies require more and more in silico biology. Grid computing supports all of these steps and, moreover, can also contribute significantly to the monitoring of ground studies to control Malaria and to the clinical tests in plagued areas.

A particularly computing intensive step in the drug discovery process is virtual screening which is about selecting in silico the best candidate drugs acting on a given target protein. Screening can be done in vitro using real chemical compounds, but this is a very expensive not necessarily error-free undertaking. If it could be done in silico in a reliable way, one could reduce the number of molecules requiring in vitro and then in vivo testing from a few millions to a few hundreds.

The WISDOM [9,10] experiment ran on the EGEE grid production service during summer 2005 to analyze one million drug candidates against Plasmepsin, the aspartic protease of Plasmodium responsible for the initial cleavage of human haemoglobin.

## 2.1 Grid Objective

A large number of applications are already running on grid infrastructures. Even if many have passed the proof of concept level [11], only few are ready for large-scale production with experimental data. Large Hadron Collider experiments at CERN, like the ATLAS collaboration [12], have been the first to test a large data production system on grid infrastructures [13]. In a similar way, WISDOM aimed at deploying a scalable, CPU consuming application generating large data flows to test the grid infrastructure, operation and services in very stressing conditions.

Docking is typically an embarrassingly parallel application, with repetitive and independent calculations. Large resources are needed in order to test a family of targets, a significant amount of possible drug candidates and different virtual screening tools with different parameter / scoring settings. This is both a computational and data challenge problem to distribute millions of docking comparisons with millions of small compound files.

Moreover, docking is the only application for distributed computing that has prompted the uptake of grid technology in the pharmaceutical industry. The WISDOM scientific results are also a means of making a demonstration of the EGEE grid-computing infrastructure for the end users community, of illustrating the usefulness of a scientifically targeted Virtual Organization, and of fostering an uptake of grid technologies in this scientific area.

## 2.2 Biological objective

Malaria is a dreadful disease caused by a protozoan parasite, plasmodium. A new strategy to fight Malaria investigated within WISDOM aims at the haemoglobin metabolism, which is one of the key metabolic processes for the survival of the parasite. Plasmepsin, the aspartic protease of Plasmodium, is responsible for the initial cleavage of human haemoglobin and later followed by other proteases [14]. It is an ideal target for rational drug design against Malaria.

Docking is a first step for in silico virtual screening. Basically, protein-ligand docking is about estimating the binding energy of a protein target to a library of potential drugs using a scoring algorithm. The target is typically a protein, which plays a pivotal role in a pathological process. The goal is to identify which molecules could dock on the protein active sites in order to inhibit its action and therefore interfere with the molecular processes essential for the pathogen. Libraries of compound 3D structures are made openly available by chemistry companies, which can produce them. Many docking software products are available either open-source or licensed.

## 3 In silico docking against avian flu

The first large scale docking experiment focused on virtual screening for neglected diseases but new perspectives appear also for using grids to address emerging diseases. While the grid added value for neglected diseases is related to their cost effectiveness as compared to in vitro testing, grids are also extremely relevant when time becomes a critical factor. A collaboration of Asian and European laboratories has analyzed 300,000 possible drug compounds against the avian flu virus H5N1 using the EGEE Grid infrastructure in April and May 2006 [15]. The goal was to find potential compounds that can inhibit the activities of an enzyme on the surface of the influenza virus, the so-called neuraminidase, subtype N1.

## 3.1 Grid Objective

Beside the biological goal of reducing time and cost of the initial investment on structure-based drug design, there are two Grid technology objectives for this activity: one is to improve the performance of the in silico high-throughput screening environment based on what has been learnt in the previous challenge against Malaria; the other is to test another environment which enables users to have efficient and interactive control of the massive molecular dockings on the Grid. Therefore, two Grid tools were used in parallel in the second data challenge. An enhanced version of WISDOM high-throughput workflow was designed to achieve the first goal and a lightweight framework called DIANE [reference] was introduced to carry a significant fraction of the deployment for implementing and testing the new scenario.

## 3.2 Biological objective

The potential for re-emergence of influenza pandemics has been a great threat since the report that avian influenza A virus (H5N1) could acquire the ability to be transmitted to humans. Indeed, an increase of transmission incidents suggests a risk of human-to-human transmission [16]. In addition, the report of the development of drug resistant variants [17] is of potential concern. Two of the present drugs (oseltamivir and zanamivir) were discovered through structure-based drug design targeting influenza neuraminidase (NA), a viral enzyme that cleaves terminal sialic acid residues from glycoconjugates. The action of NA is essential for virus proliferation and infectivity; therefore, blocking its activity generates antivirus effects. To minimize non-productive trial-and-error approaches and to accelerate the discovery of novel potent inhibitors, medical chemists take advantage of modelled NA variant structures and structure-based design. The computational tools for the work are based on molecular docking engines to carry out a quick conformational search for small compounds in the binding sites, fast calculation of binding energies of possible binding poses, prompt selection for the probable binding modes, and precise ranking and filtering for good binders.

## 4 The grid tools

### 4.1 The grid infrastructures

Compared to WISDOM which used only the EGEE infrastructure, the Avian Flu computing challenge used three infrastructures which are sharing the same middleware (LCG2) and also common services: Auvergrid, EGEE and TWGrid.

Auvergrid (http://www.auvergrid.fr/) is regional grid deployed in the French region Auvergne. Its goal is to explore how a grid can provide the resources needed for public and private research at a regional level. With more than 800 CPUs available at 12 sites, Auvergrid hosts a variety of scientific applications from particle physics to life science, environment and chemistry.

TWGrid (Taiwan Grid, http://twgrid.org/) is responsible for operating a Grid Operation Centre in Asia-Pacific region. Apart from supporting the worldwide Grid collaboration in high-energy physics, TWGrid is also in charge of federating and coordinating regional Grid resources to promote the Grid technology to the e-Science activities (e.g. life science, atmospheric science, digital archive, etc.) in Asia.

The EGEE project [3] (Enabling Grid for E-sciencE, http://www.eu-egee.org/) brings together experts from over 27 countries with the common aim of building on recent advances in Grid technology and developing a service Grid infrastructure which is available to scientists 24 hours a day. The project aims to provide researchers in academia and industry with access to major computing resources, independent of their geographic location. The EGEE infrastructure is now a production grid with a large number of applications installed and used on the available resources. The infrastructure involves more than 180 sites spread in Europe, America and Asia. The two applications described in this paper were deployed within the framework of the biomedical Virtual Organization. Resource nodes available for biomedical applications scale up to 3,000 CPUs and 21 TB disk space.

### 4.2 The WISDOM production environment

A large-scale deployment requires the development of an environment for job submission and output data collection. A number of issues need to be addressed to achieve significant acceleration from the grid deployment:
- Grid performances are impacted by the amount of data moved around at job submission. Consequently, the files providing the 3D structure of targets and compounds should preferably be stored on grid storage elements in preparation for the data challenge.
- The rate at which jobs are submitted to the grid resource brokers must be carefully monitored in order to avoid their overload. The job submission scheme must take into account this present limitation of the EGEE brokering system.
- Grid submission process introduces significant delays for instance at the level of resource brokering. The jobs submitted to the grid computing nodes must be sufficiently long in order to reduce the impact of this middleware overhead.

The WISDOM production environment was designed to achieve production of a large amount of data in a limited time using EGEE, Auvergrid and TWGrid middleware services. Three packages were developed in Perl and Java. Their entry points are a simple command line tool. The first package installs the application components (software, compounds database…) on the grid computing nodes. The second package tests these components. The third package monitors the submission and the execution of the WISDOM jobs. Figure 1 illustrates the packages and the steps of the jobs execution.

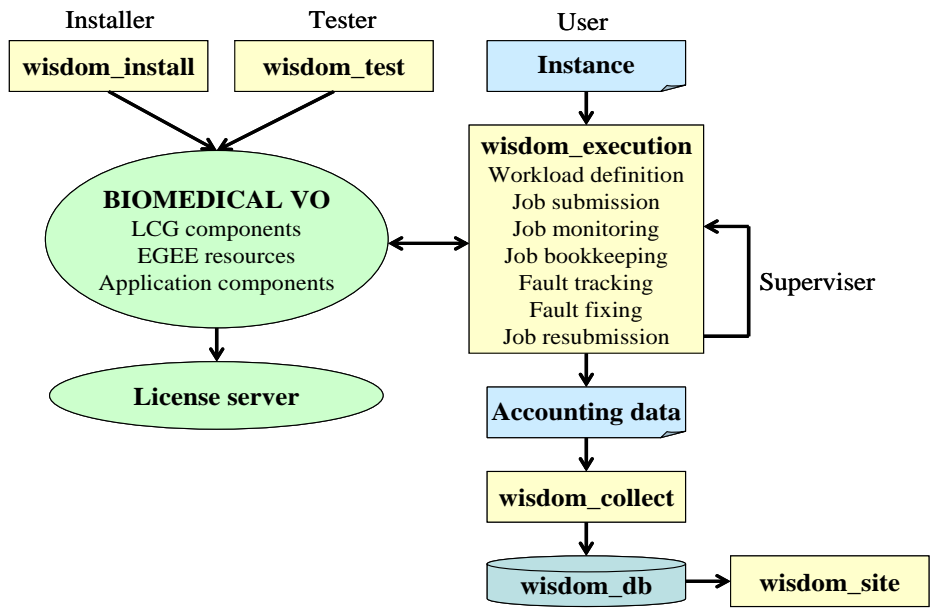

**Fig. 1.** Design of the WISDOM production system

The environment was improved to address limitations and bottlenecks identified during the first data challenge against Malaria deployed in the summer of 2005 on the EGEE infrastructure. For instance, the number of resource broker machines and the rate at which the jobs were submitted to these grid services were reduced to avoid their overload. Another improvement concerned the resubmission process after a job failure, which was redesigned to avoid "sink-hole" effect on a failing grid-computing node. Automatic resubmission was replaced by the manual intervention of the WISDOM production user.

### 4.3 The DIANE framework

DIANE (Distributed Analysis Environment, http://cern.ch/diane/) is a lightweight distributed framework for parallel scientific applications in master-worker model. It assumes that a job may be split into a number of independent tasks, which is a typical case in many scientific applications. It has been successfully applied in a number of applications ranging from image rendering to data analysis in high-energy physics.

As opposed to standard message passing libraries such as MPI [18], the DIANE framework takes care of all synchronization, communication and workflow management details on behalf of the application. The execution of a job is fully controlled by the framework, which decides when and where the tasks are executed. Thus the application is very simple to program and contains only the essential code directly related to the application itself without bothering about networking details. Aiming to efficiently bridge underlying distributed computing environments and application centric user interface, DIANE itself is a thin software layer which can easily work on top of more fundamental middleware such as LSF, PBS or the Grid Resource Brokers. It may also work in a standalone mode and does not require any complex underlying software.

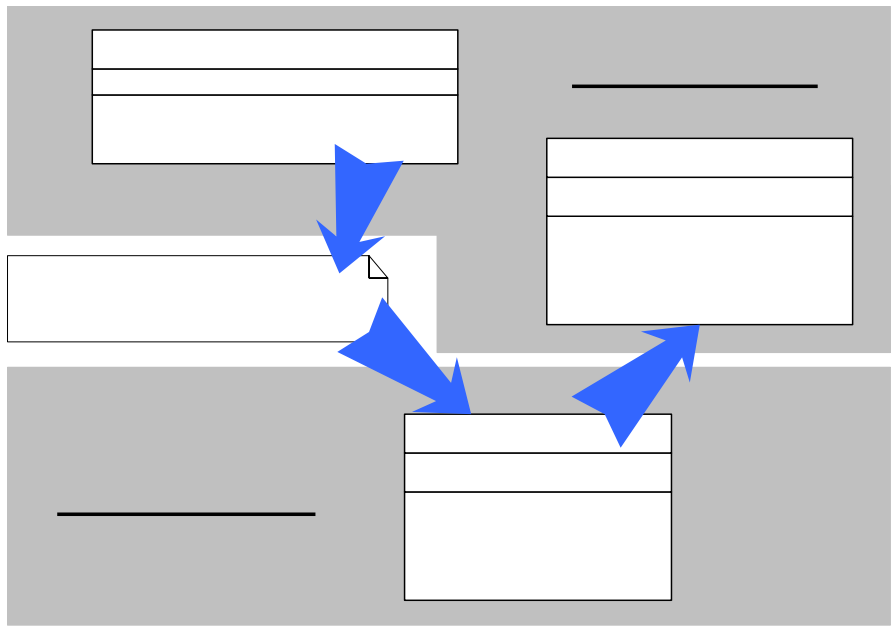

**Fig. 2.** Template of DIANE application plug-ins

As a framework, DIANE provides an adapter for applications. Figure 2 shows the template of DIANE application plug-ins. A complete DIANE application plug-in should implement three major Python objects: the Planner and the Integrator objects implement the job splitting and result merging, respectively; while the logic of the Worker object concentrates on the execution of the individual task. When a DIANE job is started by a user, both the Planner and the Integrator objects are invoked by a master agent usually executed on the user's desktop, and typically the worker agents are submitted to run on distributed CPUs such as the Grid worker nodes.

Once the worker agent is launched, it first registers itself with the master agent. In the second step, a channel is established for pulling the tasks from the queue held by the master agent. When the worker agent has finished the individual task, the result is returned and merged on the master. The pulling-executing-returning cycle will iterate until all the tasks are accomplished. The same channel is also used to profile the worker agent's health and to support user interaction with the task. The whole DIANE framework is written in Python and the communication between the master agent and the worker agents is based on the CORBA protocol.

# 5  Results

## 5.1  WISDOM results

The WISDOM experiment ran on the EGEE grid production service from 11 July 2005 until 19 August 2005. It saw over 46 million docked ligands, the equivalent of 80 years on a single PC, in about 6 weeks. Up to 1,700 computers were simultaneously used in 15 countries around the world. Commercial software with a server license was successfully deployed on more than 1,000 machines at the same time. WISDOM demonstrated how grid computing can help drug discovery research by speeding up the whole process and reduce the cost to develop new drugs to treat diseases such as Malaria.

**Grid performances**
The data challenge was a very useful experience to identify the limitations and bottlenecks of the EGEE infrastructure. The overall grid efficiency was on average about 50%. This means that a large fraction of the jobs had to be resubmitted. This generated a significant extra workload for the users.

The different sources of failures are identified on table 1 with their corresponding rates.

In the phase where licensed software was deployed on the grid, the dominant origin of failure was the license server in charge of distributing tokens to all the jobs running on the grid. The development of a grid service to manage license software is under way to address this single point of failure ignored by the information system.

The second most important sources of failure were workload management and site failures, including overload, disk failure, node mis-configuration, disk space problem, air-conditioning and electrical cut. To improve the job submission process, an automatic resubmission of jobs was included in the WISDOM execution environment. However, the consequence of automatic resubmission was the creation of several "sinkhole" effects where all the jobs were attracted to a single node. These sinkhole effects were observed when the status of a Computing Element was not correctly described in the information system. If a Computing Element already loaded is still viewed as completely free by the Information System, it keeps receiving jobs from the Resource Broker. If the Computing Element gets down, all jobs are aborted. Even if the Computing Element can support the excessive number of jobs, the processing time is going to be very long.

**Table 1.** Origin of failures during the WISDOM deployment with their corresponding rates

|  | Rate | Reasons |
|---|---|---|
| Success rate after checking output data | 46 % |  |
| Workload Management failure | 10 % | Overload, Disk failure |
|  |  | Mis-configuration, disk space problem |
|  |  | Air-conditioning, power cut |
| Data Management failure | 4 % | Network / connection |
|  |  | Power cut |
|  |  | Other unknown causes |
| Sites failure | 9 % | Mis-configuration, Tar command, disk space |
|  |  | Information system update |
|  |  | Job number limitation in the waiting queue |
|  |  | Air-conditioning, Electrical cut |
| Unclassified | 4 % | Lost jobs |
|  |  | Other unknown causes |
| Server license failure | 23 % | Server failure |
|  |  | Power cut |
|  |  | Server stop |
| WISDOM failure | 4 % | Job distribution |
|  |  | Human error |
|  |  | Script failure |

The WISDOM production system developed to submit the jobs on the grid accounted for a small fraction of the failures, as did also the grid data management system. About 1TB of data were produced by the 72,000 jobs submitted. Collection and registration of these output data turned out to be a heavy task. The grid data management services allowed replicating all the output files for backup. However, they did not allow storing all the results directly in a database.

Finally, unclassified failures accounted for 4% inefficiency. This illustrates the work, which is still needed to improve grid monitoring.

In terms of grid throughput, the resource brokers significantly limited the rate at which the jobs could be submitted. Another significant source of inefficiency came from the difficulty for the grid information system to provide all the relevant information to the resource brokers when they distribute the jobs on the grid. Therefore, job scheduling was a time-consuming task for the WISDOM users during all the data challenge due to the encountered limitations of the information system, the computing elements and the resource brokers. The achieved throughput averaged about 12 docked ligands per second during the whole computing challenge.

Several of the issues identified during WISDOM deployment were improved for the second large scale docking targeted on avian flu which is described in the next chapter.

### 5.2 Results from avian flu computing challenge

Table 2 summarizes the achieved deployments using WISDOM and DIANE environments.

**Table 2.** Statistical summary of the WISDOM and DIANE activities

|  | WISDOM | DIANE |
|---|---|---|
| Total number of completed dockings | 2,160,095 | 308,585 |
| Estimated duration on 1 CPU | 88.3 years | 16.7 years |
| Duration of the experience | 6 weeks | 4 weeks |
| Cumulative number of Grid jobs | 54,000 | 2,585 |
| Maximum number of concurrent CPUs | 1,700 | 240 |
| Number of used Computing Elements | 69 | 36 |
| Crunching factor | 767 | 203 |
| Approximated distribution efficiency | 46% | 84% |

During the avian flu computing challenge, the WISDOM production environment was used to distribute 54,000 jobs on 69 Grid Computing Elements to dock about 2 million pairs of target and chemical compound for a total amount of about 88 CPU years. Because the Grid resources were used by other virtual organizations during the data challenge, a maximum of 1,700 CPUs were concurrently running at the same time. For the DIANE part, we were able to complete 308,585 docking runs (i.e. 1/8 of the whole challenge) in 30 days using the computing resources of 36 Grid nodes. A total number of 2,580 DIANE worker agents have been running as Grid jobs during that the DIANE master concurrently maintained period and 240 of them. About 750 GBytes of date have been produced on the Grid during the data challenge.

**Grid performances**
Since a Grid is a dynamic system in which the status of resources may change without central control, transient problems occur which cause job failures. In the WISDOM activity, about 83% of the jobs were reported as successfully finished according to the status logged in the Grid Logging and Bookkeeping system; the observed failures were mainly due to errors at job scheduling time because of mis-configuration of Grid Computing Elements. However, the success rate went down to 80% after checking the content of the data output file and subtracting human failures. The main cause for these failures was frequent last-minute error in the transfer of results to the Grid Storage Elements. Compared to the previous data challenge, improvement is significant as the observed success rates were respectively 77 and 63%. The last-minute error in output data transfer is particularly expensive since the results are no longer available on the Grid Worker Node although they might have been successfully produced.

In DIANE, a similar job failure rate was also observed; nevertheless, the failure recovery mechanism in DIANE automated the re-submission and guaranteed a finally fully complete job. On the other hand, the feature of interactively returning part of the computing efforts during the runtime (e.g. the output of each docking) also introduces a more economical way in using the Grid resources.

For the instances submitted using WISDOM production environment, the overall crunching factor was about 767. The corresponding distribution efficiency defined as the ratio between the overall crunching factor and the maximum number of concurrently running CPUs was estimated to 46%. This is due to the known issue of long job waiting time in current EGEE production system.

The task pull model adopted by DIANE allows isolating the scheduling overhead of the Grid jobs and is therefore expected to achieve a better distribution efficiency. During the data challenge, DIANE was able to push the efficiency to higher than 80% within the scope of handling intermediate scale of distributed docking. A fare comparison with WISDOM can only be made if the improvement of the DIANE framework is tested in a large scale as the exercise of WISDOM. Because of the highly scalable nature of the WISDOM framework, high throughput docking could be achieved at a rate of 2 seconds per docking. As DIANE was handling not more than a few hundred concurrent jobs, its throughput was limited to about one docking every 9 seconds. The difference observed between the throughputs achieved during the WISDOM and avian flu computing challenges is because each docking computation was very significantly longer during the avian flu data challenge.

## 6    Perspectives

The results obtained by the first two virtual docking computing challenges have opened important perspectives. On the grid side, developments are under way to further improve the WISDOM and DIANE environments to improve the quality of service offered to the end users. From an application point of view, beyond the necessity to analyze the results obtained on Malaria and avian flu, it is particularly relevant to further enable the deployment of virtual screening on large-scale grids.

### 6.1    Perspective on the grid side: WISDOM-II

The impact of the first WISDOM computing challenge has significantly raised the interest of the research community on neglected diseases so that several laboratories all around the world have expressed interest to propose targets for a second computing challenge called WISDOM-II. Contacts have been established with research groups from Italy, United Kingdom, Venezuela, South Africa and Thailand and discussions are under way to finalize the list of targets to be docked.

As well, several grid projects have expressed interest to contribute to the WISDOM initiative by providing computing resources (Auvergrid, EELA (http://www.eu-eela.org/), Southeast Asia Grid, Swiss Biogrid (http://www.swissbiogrid.com/)) or by contributing to the development of the virtual screening pipeline (Embrace (http://www.embracegrid.info/), BioinfoGRID (http://www.bioinfogrid.eu/)). Of particular interest for WISDOM-II is the deployment of a common application on several infrastructures using different middlewares like EGEE and Swiss Biogrid.

### 6.2    From virtual docking to virtual screening

While docking methods have been significantly improved in the last years, it is general opinion that docking results need to be post-processed with more accurate modelling tools before biological tests are undertaken, such as Molecular dynamics

(MD). However, for the same number of compounds, MD analysis requires much heavier computing than docking. Consequently, MD can only be applied to a restricted number of compounds, usually the best hits coming out of the docking step. MD and subsequent free-energy analysis most often changes significantly the scoring of the best compounds and it is therefore very important to apply it to as many compounds as possible. Grids appear very promising to improve the virtual screening process by increasing the number of compounds that will be processed using MD.

For instance, running MD analysis with Amber 8 software on 1 protein target and 10,000 compounds requires about 50 CPU years for a model with few thousands atoms and few 10,000s of steps. This process produces only 10s GB of data. One run with only one compound takes 44 hours (computing time heavily depends on choice of conditions, like explicit water simulations, or generalized born simulations).

Both grids of clusters such as EGEE and grids of supercomputers such as DEISA (http://www.deisa.org/) are relevant for MD computing. MD computations of large molecules are significantly faster on a supercomputer. Within the framework of the BioinfoGRID European project, focus will be put on the reranking of the best scoring ligands coming out of WISDOM. The goal will be to deploy on at least one of the European grid infrastructures MD software to rerank the best compounds before in vitro testing.

## 7    Conclusion

Large-scale grids offer unprecedented perspectives for in silico drug discovery. This paper has presented pioneering activities in the field of grid enabled virtual screening against neglected and emerging diseases in Europe. This activity started with a large scale docking deployment against Malaria on the EGEE infrastructure in 2005, which was followed by another computing challenge focused on avian flu in the spring of 2006. A second computing challenge on neglected diseases is foreseen to take place in the fall: it will involve several European projects, which bring additional resources and complementary contributions in order to enable the complete virtual screening process on the grid.

These deployments were achieved using two different systems for submission and monitoring of virtual docking jobs. The WISDOM system was designed to achieve production of a large amount of data in a limited time while the DIANE framework was designed as a lightweight distributed framework for parallel scientific applications in master-worker model. Both systems were able to provide high throughput virtual docking. The DIANE framework demonstrated better performances in terms of job success while the WISDOM system allowed to achieve higher throughput. Development of a new system merging functionalities from both WISDOM and DIANE frameworks is under way in the perspective of WISDOM-II, second computing challenge on neglected diseases.

**Acknowledgment.** This work is co-funded by the European Commission through the EGEE and BioinfoGRID projects and the Embrace network of excellence. The authors express particular thanks to the site managers in EGEE, TWGrid, AuverGrid for operational supports, the LCG ARDA group for the technical support of DIANE, and the Biomedical Task Force for its participation to the WISDOM deployment. The following institutes contributed computing resources: ASGC (Taiwan); NGO (Singapore); IPP-BAS, IMBM-BAS, IPP-ISTF (Bulgaria); CYFRONET (Poland); ICI (Romania); CEA-DAPNIA, CGG, IN2P3-CC, IN2P3-LAL, IN2P3-LAPP, IN2P3-LPC (France); SCAI (Germany); INFN (Italy); NIKHEF, SARA, VL-e (Netherlands); IMPB RAS (Russia); UCY (Cyprus); AUTH FORTH-ICS, HELLASGRID (Greece); RBI (Croatia); TAU (Israel); CESGA, CIEMAT, CNB-UAM, IFCA, INTA, PIC, UPV-GryCAP (Spain); BHAM, University of Bristol, IC, Lancaster University, MANHEP, University of Oxford, RAL, University of Glasgow (United Kingdom).